\documentclass[sigconf, nonacm]{acmart}
\usepackage[utf8]{inputenc}
\usepackage[english]{babel}
\usepackage{lmodern}
\usepackage[T1]{fontenc}
\usepackage{algorithmic}
\usepackage{graphicx}
  \graphicspath{{images/}}
\usepackage{subcaption} 
\usepackage{tabularx}
\usepackage{textcomp}
\usepackage{xcolor}
    
 \AtBeginDocument{%
  \providecommand\BibTeX{{%
    \normalfont B\kern-0.5em{\scshape i\kern-0.25em b}\kern-0.8em\TeX}}}

\newcommand{\eg}{e.\,g.\ }

\begin{document}

\title{Secure and Efficient Tunneling of MACsec for Modern Industrial Use Cases}

\author{Tim Lackorzynski}
\affiliation{%
  \institution{TU Dresden}
  \city{Dresden}
  \country{Germany}
}
\email{tim.lackorzynski@tu-dresden.de}

\author{Sebastian Rehms}
\affiliation{%
  \institution{TU Dresden}
  \city{Dresden}
  \country{Germany}
}
\email{sebastian.rehms@tu-dresden.de}

\author{Tao Li}
\affiliation{%
  \institution{TU Dresden}
  \city{Dresden}
  \country{Germany}
}
\email{tao.li@tu-dresden.de}

\author{Stefan K\"{o}psell}
\affiliation{%
  \institution{Barkhausen Institute / CeTI TU Dresden}
  \city{Dresden}
  \country{Germany}
}
\email{stefan.koepsell@barkhauseninstitut.org}

\author{Hermann H\"{a}rtig}
\affiliation{%
  \institution{TU Dresden}
  \city{Dresden}
  \country{Germany}
}
\email{hermann.haertig@tu-dresden.de}

\renewcommand{\shortauthors}{Lackorzynski and Rehms, et al.}

\begin{abstract}
Trends like Industry 4.0 will pose new challenges for future industrial networks.
Greater interconnectedness, higher data volumes as well as new requirements for speeds as well as security will make new approaches necessary.
Performance optimized networking techniques will be demanded to implement new use cases, like network separation and isolation, in a secure fashion.

A new and highly efficient protocol, that will be vital for that purpose, is MACsec.
It is a Layer 2 encryption protocol that was previously extended specifically for industrial environments.
Yet, it lacks the ability to bridge local networks.

Therefore, in this work, we propose a secure and efficient Layer 3 tunneling scheme for MACsec.
We design and implement two approaches, that are equally secure and considerably outperform comparable state-of-the-art technique techniques.
\end{abstract}


\maketitle

\section{Introduction}

Industrial networks currently find themselves in a phase of change. New use cases are being implemented along trends that are variably subsumed under the terms Industry 4.0 or the Industrial Internet of Things.
At the same time components, like industrial machines, have very long lifetimes and high investment costs. These legacy devices will stay and must be integrated into networks of the future.
Many heterogeneous devices and cloud-based software components from different vendors will be present at the same time.
The concept of perimeter security applied to formerly isolated factory networks will not offer sufficient protection any longer.
Hence, factory networks have to be viewed as zero trust networks.

As updating or replacing machinery is often not possible, these legacy devices must be integrated.
Techniques like network isolation and encryption will be used to reduce their attack surface and to protect the data traffic that must flow over insecure networks.
Legacy industrial machines will be protected via so-called industrial encryption gateways, like \cite{fastvpn}.
These physical devices are put in front of an industrial machine and transparently encrypt the traffic data before admitting the now protected data to the network.
Associated gateways can then decrypt that traffic and hence realize a secure connection over an insecure network.

A promising protocol for that purpose is MACsec \cite{macsec}.
It is an efficient and standardized Layer 2 encryption protocol available in the Linux kernel.
Hence, it can readily be applied to the resource-restricted embedded platforms prevalent in industrial ecosystems.
Previous works identified MACsec as the best choice for encryption gateways \cite{my_vpn_study} and already extended it for the industrial environment \cite{my_macsecpp}.
Yet, one drawback still remains. 
MACSec works purely on Layer 2.
While this allows to use MACsec to protect a wide variety of established industrial protocols, it also restricts its use to one Local Area Network (LAN).
Therefore, we want to extend MACsec with the ability to tunnel or bridge its traffic between multiple local networks.

The state-of-the-art approach of tunneling over public networks is to use a Virtual Private Network (VPN) protocol.
Yet, it has been shown, that this reduces the performance considerably in encryption gateway use cases \cite{my_vpn_study}.
Fully re-encrypting MACsec frames is also not actually necessary, as the payload is already encrypted by MACsec.
This results in additional overheads without delivering any additional benefit.

Following from that, the straightforward and naive approach would be to use a non-encrypting tunneling protocol like L2TP \cite{rfc3931}, VXLAN \cite{rfc7348} or GRE \cite{rfc1701} and only rely on the security properties of MACsec.
This, however, has two main drawbacks: a lack of confidentiality of the frame headers and missing authentication at the tunnel end points.
Both will be discussed in detail below.
We want to extend the naive approach by providing the missing properties and by focusing on the headers.

In this work, we want to propose two different strategies on how to protect these headers such that a MACsec frame can be securely tunneled over insecure networks without touching the already encrypted payload within that frame.
The rationale hereby is, that by only working on the comparatively small headers, big increases in efficiency can be gained, compared to the state-of-the-art, that just encrypts the whole frame as is.
The two approaches investigated are equally secure and differ in terms of complexity.
Both will heavily focus on improving performance as much as possible.

The remainder of this work is structured as follows.
Sec.~\ref{sec:background} first introduces the header structure of MACsec in detail, including how sensitive the individual fields inside the header are towards attackers. Secondly, as efficiency is a key factor in our investigations, it will discuss the topic of fast packet processing as a means to optimize the performance of our approaches.
Sec.~\ref{sec:related_work} will discuss related work.
Sec.~\ref{sec:design} presents the two approaches and Sec.~\ref{sec:implementation} details our implementation.
Results are discussed in Sec.~\ref{sec:evaluation}.
Sec.~\ref{sec:conclusion} gives concluding remarks.
  
\section{Background}
\label{sec:background}

\begin{figure}
  \begin{center}
    \includegraphics[scale=1.0,keepaspectratio=true]{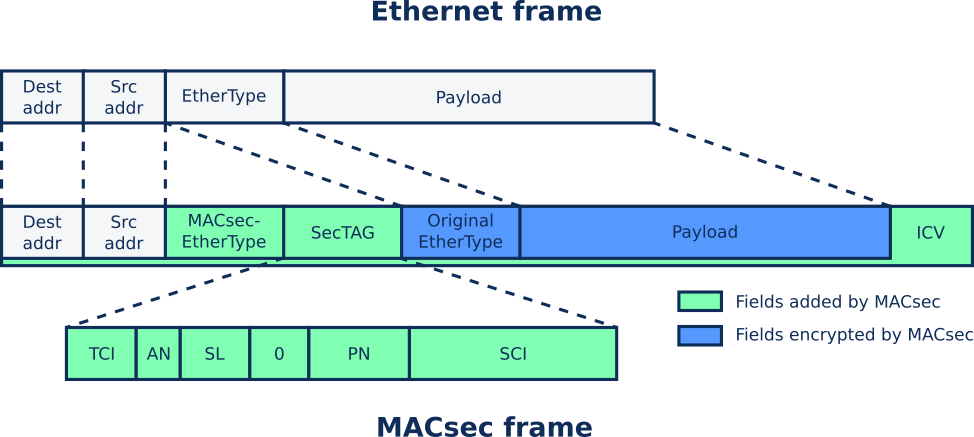}
    \caption{MACsec applied to unprotected MACsec frame.}
  \label{fig:frame_structure_eth_macsec}
  \end{center}
\end{figure}

MACsec protects Layer 2 Ethernet frames by encrypting the frames' payload and by integrity protecting the whole frame, as shown in Fig.~\ref{fig:frame_structure_eth_macsec}.
It adds at maximum 32 byte of additional header that get integrity protected as well.
Destination and source addresses of the original frame are kept.
The original EtherType is moved to the payload (data segment) and the MACsec exclusive EtherType is set instead.
MACsec is an ideal protocol for encryption gateways that transparently protect the data traffic of legacy industrial machines as it encapsulates whole Ethernet frames independently of any higher layer protocol.

As the figure shows, the payload of the MACsec frame is already encrypted.
Yet, the MACsec header fields remain readable.
To be able to tunnel these frames in a smarter way compared to the initial approaches discussed above, we will now investigate in detail, which of the header fields is security sensitive and needs protection.

MACsec defines communication relationships as Secure Channels (SC).
These are configured unidirectional one-to-many, meaning that a sender can send data to multiple receivers.
A corresponding SC has to be configured on all receivers and when two MACsec entities want to communicate to each other, two SCs must be configured where both are respectively sender and receiver.
A SC is identified by the 64 bit Secure Channel Identifier (SCI).
It consists of the 48 bit MAC address of the MACsec device plus a 16 bit port number.
The configuration of ports allows to have multiple MACsec instances on the same device.
The SCI can also be omitted if sender and receiver are directly connected.
This reduces the overall MACsec header size, but it also is a special case that requires direct cabling between MACsec devices. This case is ignored here.

The 32 bit Packet Number (PN) is used to establish an order of sent frames and also acts as the counter for the AES-GCM cipher that is used for encryption.
Galois Counter Mode (GCM) dictates that for security reasons a key may never be used with the same counter (PN) twice.
Therefore, when the PN overflows, the old encryption key is discarded and a new key is selected.
These keys are hence ephemeral. 

MACsec allows for a certain amount of asynchrony within a defined replay window.
To guarantee this window even when the PN overflows, the concept of Security Associations (SAs) is introduced.
The actual encryption keys are not bound to an SC but to a security association.
An SA is defined by the SCI plus a 2 bit Association Number (AN).
When the PN reaches its maximum value, the next AN is chosen, resulting in a reset of the PN to 0 and a different encryption key being selected.
To allow for smooth rollover two ANs can be active at the same time and a receiver can choose which key to use for decryption based on the AN number found in the header.
Keys can either be configured by hand for a specific SA or be provided by the MACsec Key Agreement protocol (MKA) that manages MACsec stations automatically \cite{mka}.

The 6 bit Short Length (SL) field is set to zero if the payload is larger than 48 bytes and represents the length of the payload otherwise.

The 6 bit Tag Control Information (TCI) field is comprised of different single bit flags.
The Version (V) bit is always (as per the standard) set to 0, while the End Station (ES) bit indicates that the sending MACsec station is also the source of the frame.
In effect a set ES flag means that the frames' Ethernet source address is identical to the first 48 bits of the SCI.
The SCI present (SC) flag shows presence of the SCI. As described above, there might be cases, when the SCI can be omitted.
The Single Copy Broadcast (SCB) flag is used in fiber tree topologies to toggle single copy broadcasts without the explicit use of the SCI.
This scenario is also not relevant to this work.
The Encrypted payload (E) flag indicates that the payload is additionally encrypted instead of only being integrity-protected.
In this work, we always assume that this is the case.
The Changed text (C) flag indicates whether the data segment is simple payload or management information addressed to the MKA client running on the station.
Two bits (0) exist in the header that are as of yet undefined in the MACsec standard and therefore are always set to 0. They are reserved for future use.
A 16 bit Integrity Check Value (ICV) is calculated over the whole frame and is appended at the end.
This value provides integrity protection for all headers and the payload of the frame, irrespective if they originate from the original frame or additions by MACsec.
This value is always calculated and, in contrast to the payload encryption, not optional.

After this short introduction of the MACsec header fields, we will now look at them individually on how sensitive towards security they are.
Specifically, what information could an attacker derive when observing the headers when they are tunneled in an unprotected fashion, as described with our "naive" tunneling approach discussed above.

First, an attacker would see the Ethernet source and destination addresses of the tunneled frame.
These are the addresses of the original machines that needed protection in the first place.
This would allow the attacker to attribute communication partners and relationships as well as data flows between them.
He would also learn about the traffic patterns of the communication, including volume and frequency.
Additionally, Ethernet addresses (MAC addresses) typically include information about the vendor of the network card. This might give clues to identify the specific industrial machine or type of machine that is communicating.

An attacker would also learn the EtherType of the tunneled frame (MACsec) and hence would learn how to interpret the following byte fields (that may otherwise look random).

The SCI allows to identify the MACsec devices (the encryption gateways). This would give insights on the structure of the internal network by showing which industrial machines are grouped behind the same gateway.
Recording observed SAs and PNs would reveal traffic patterns of the communication. These values on themselves would also make it possible to discern individual flows.

In contrast, the SL field does not offer information to the attacker that he cannot learn anyhow by just observing the length of the frames.
The TCI field also does not leak any new information, as the attacker can infer flags anyway based on what is observable.
The only interesting flag is the C flag, that determines whether ordinary payload or MKA traffic is transmitted.
As MKA traffic will be handled differently in our design, it can be ignored here.
The payload, including the original EtherType, is encrypted anyhow, so the attacker cannot learn anything, except the length.
The ICV is a cryptographically strong Message Authentication Code (MAC), that is for the attacker not discernible from mere random bits. He especially is not able to use this value to create bogus frames that would pass the integrity check or derive information about the encryption key from it.
Fig.~\ref{fig:macsec_sensitive} summarizes the above and depicts which header fields of a MACsec frame we consider sensitive.
\\

MACsec-tunneling will be facilitated by a single device that will sit at the edge of the local network.
Hence, it must handle the traffic flows of all encryption gateways in that domain in parallel.
And as network performance demands in general will only grow in future industrial networks, we also aim to be as efficient as possible.
Therefore, we will now focus on the topic of fast packet processing.

In contemporary general purpose systems, the performance is limited by the architecture of the network software stack and not by the data rates of the physical network interfaces \cite{fast_packet_processing}.
Additionally, most transmission overheads typically stem from per-packet processing steps.
This means that the size of a packet or frame has only a secondary influence on the processing times.
It additionally means, that small packets, that are prevalent in industrial scenarios, can also profit greatly from optimizations from that direction. 

The networking performance in general can be enhanced by various software- and hardware-based approaches.
Hardware-based approaches include network cards that are equipped with special Field-Programmable Gate Arrays (FPGAs), so-called SmartNICs (Smart Network Interface Controller).
These are special integrated circuits that can be directly programmed by the user. These programs run directly on hardware, increasing their performance considerably, compared to implementations running on-top of the operating system.
Yet, these approaches depend on specific hardware extensions as well as interfacing software that integrates these features with the general purpose operating system that is still needed.
Additionally, these techniques are, as of yet, only found in cloud computing environments and not in the sphere of industrial computing.

Therefore, we focussed on a software-based approach instead.
There are different software frameworks available that implement various techniques that optimize networking performance, like for example reducing kernel interrupts and system calls, zero-copy and memory mapping or batch processing and parallelism.

The most prominent frameworks for this purpose are netmap, PF\_RING\_ZC and DPDK.
According to \cite{io_frameworks}, DPDK shows the best performance of the three, while PF\_Ring is a close second. It additionally provides better access to hardware features and offloading.
Therefore, we based our design on DPDK.

The Data Plane Development Kit (DPDK)\footnote{https://www.dpdk.org/} provides an extensive software library for high-speed packet processing and tightly integrates with other functions that are useful in this context, like hash table management.
And while DPDK is mainly supported by and traditionally focussed on NICs from Intel, support for network controllers from other vendors is steadily increasing\footnote{https://core.dpdk.org/supported/}.
After loading a DPDK driver into the system and binding it to a specific NIC, the network controller is then invisible to the kernel and hence outside the standard Linux kernel stack.
This makes it easily possible to dedicate the physical device to one use case and not suffer from overheads introduced by secondary services provided by the general purpose operating system.

\section{Related Work}
\label{sec:related_work}

The use case of bridging or tunneling networks, so that participants can connect to each other as if they were in the same network, is well researched.
The L2 Tunneling Protocol (L2TPv3) is a standard tool for that purpose, that allows to transmit Layer 2 Ethernet frames over Layer 3 IP-based networks \cite{rfc3931}.
It encapsulates the Ethernet frames into UDP packets, but does not protect its payload in any way.
No encryption or integrity protection takes place.
Therefore, L2TP should only be used in public networks (the Internet) together with the IPsec protocol.
IPsec adds the necessary data encryption and integrity protection.
This approach is canonically called L2TP/IPsec \cite{rfc3193}.
Other standard VPN( Virtual Private Network) protocols, that could be used likewise, are OpenVPN\footnote{https://openvpn.net/} and Wireguard \cite{wireguard}.

In contrast to our intended use case, these approaches only offer point-to-point connectivity. Multiple end points are not supported.
Virtual Extensible LAN (VXLAN) on the other hand offers such point-to-multipoint connectivity \cite{rfc7348}.
It was specifically designed for this purpose, but also does not offer any protection of the tunneled traffic data.

The bridging of industrial networks will only become a more important subject in the future, for the reasons introduced above.
Many concrete proposals have been published.
Yet, even newer approaches lack security considerations (like \cite{no_sec_gw}) and hence cannot be applied to our use case, where we assume zero trust networks.
Other works consider security but employ new technologies and disregard the need for backwards compatibility for legacy components \cite{meta_overlay}.

MACsec on the other hand was previously enhanced and optimized for the industrial environment.
What it is missing is specifically the ability to bridge networks.
Just adding a state-of-the-art VPN solution for that purpose, would incur unnecessary overheads, as already mentioned. 
Since no previous approach addresses all requirements, this work wants to add the feature of network bridging to MACsec to make MACsec a complete protocol that can be generally applied to our setting.

\section{Design}
\label{sec:design}

To successfully implement tunneling of already encrypted MACsec frames, we designed two approaches.
One is a more simple one that only encrypts the headers, while the second one is more complex, but optimized towards performance as much as possible.
In the following, we will first describe our scenario in more detail and discuss requirements as well as some further necessary concepts.
Based on that, we will first discuss the more complex identifier-based approach, which is then followed by the encryption-based approach.

\subsection{Scenario and Requirements}

\begin{figure}
  \begin{center}
    \includegraphics[scale=1.0,keepaspectratio=true]{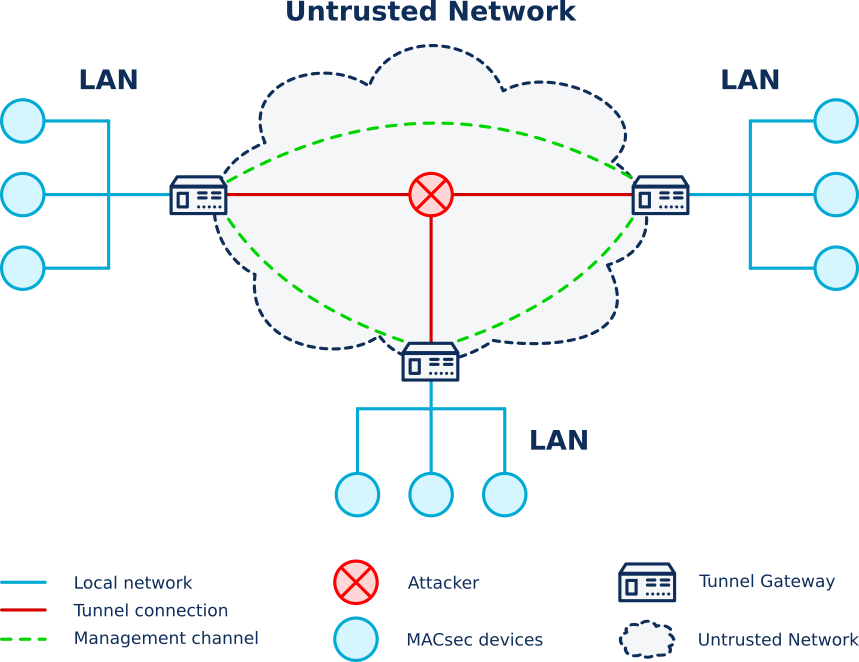}
    \caption{Local networks bridged by tunnel gateways.}
  \label{fig:scenario}
  \end{center}
\end{figure}

Fig.~\ref{fig:scenario} shows the scenario, we based our designs on.
We assume an untrusted Layer 3 IP network, which multiple Local Area Networks (LANs) are attached to. Tunnel gateways act as interfaces in between.
The LANs are populated by devices that speak MACsec.
These devices are, just as introduced above, actually encryption gateways protecting legacy industrial machinery.
Yet, for the sake of clarity, we abstract from that and just assume, they are devices that emit MACsec frames.

The basic idea of tunneling is to enable the local devices (in blue) to transparently communicate among each other, irrespective in which concrete local network both partners reside.
A remote MACsec device should appear as if it was part of the own local network.

The tunnel gateways do not act as MACsec communication partners and only facilitate the tunneling.
Only one tunnel gateway can be configured per LAN, as loops would be possible otherwise.
We assume the tunnel gateways to be preconfigured, so that they have knowledge of each other and can transmit data in between.

Tunnel gateways maintain two distinct channels. The first is the tunnel over which actual MACsec frames are transmitted (in red), while the second is a management channel (in green) that is used for the exchange of information that is necessary to synchronize the gateways.
Since the management channel is not supposed to transport performance-critical traffic, we just assume it is protected using a standard VPN protocol, for example Wireguard.
Although  MACsec can in principle be configured to not encrypt, we assume encryption is always enabled.

Based on this scenario, our designs are required to implement certain aspects.
As discussed above, the security-critical parts of a MACsec frame are its headers.
Hence, they must be protected, so that an attacker eavesdropping on the untrusted network cannot gain any information.
Second, our designs must authenticate the tunnel traffic. This means that it becomes possible for a tunnel gateway to discern between real and fake traffic including replays.
Without that requirement, an attacker could run a Denial of Service (DoS) attack on the MACsec devices inside the local networks, as the traffic would only be checked for integrity by them.
We assume the MACsec devices to be of the type of typical industrial embedded platforms.
This means, they are somewhat resource-restricted or rather on the lower end of the performance scale and hence easily overwhelmed by even moderate DoS attacks.
Furthermore, we require our designs to allow for point-to-multipoint tunneling, meaning that it is possible for MACsec devices to communicate to remote devices in more than one other LAN, just as depicted in the figure.
As a last requirement, we demand the tunnel connections that bridge the MACsec frames to be optimized for performance (in contrast to the management channel).

As already mentioned, we assume the attacker to sit on the network over which the MACsec frames are being tunneled.
The attacker can read, modify or drop packets.
We do not assume an attacker can drop all packets because this would result in completely different countermeasures and exceed the abilities of networking protocols in general.
Yet, we of course assume that some packet loss is possible, as this must not necessarily stem from a deliberate attack but from some random temporary network failure.
We also assume the attacker to be able to replay and forge bogus messages.
Yet, in our model, the attacker cannot break strong cryptographic primitives.

Finally, we will define the concept of flows more detailed, as they are necessary to understand the remainder of our designs.
The necessity to define flows in the first place stems from the fact that the MACsec frames being relayed between the tunnel gateways are not addressed from and to the MACsec devices but the actual end nodes (industrial machines) that they protect.
The MACsec devices as well as the tunnel gateways do not know the destination and hence the target LAN of any frame that it sees.
When MACsec is rolled out only inside a single LAN, this is not a problem, because all stations are part of the same shared medium or broadcast domain.
A MACsec station would see every frame and check the SCI, whether some action was necessary.
But, since we do not want to stupidly broadcast all frames between LANs, we need to define some notion of data flows, so that once a frame is identified to belong to a certain flow, it can be relayed directly, analogous to how switches learn outgoing ports to reduce network load.

We define a flow as unidirectional traffic from one MACsec device to another.
This is in contrast to MACsec itself, where a SC is defined between one sender and multiple receivers. 
We identify a flow by the SCI and SA fields in the MACsec header together with the destination MAC address.

\subsection{Identifier-based Approach}

\begin{figure}
  \begin{subfigure}{\columnwidth}
  \begin{center}
  \includegraphics[scale=1.0,keepaspectratio=true]{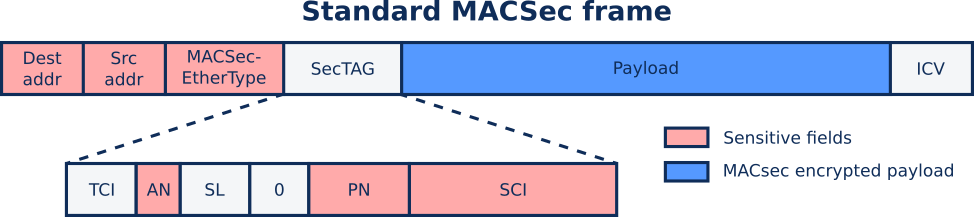}
  \caption{}
  \label{fig:macsec_sensitive}
  \end{center}
  \end{subfigure}
  \begin{subfigure}{\columnwidth}
  \begin{center}
  \includegraphics[scale=1.0,keepaspectratio=true]{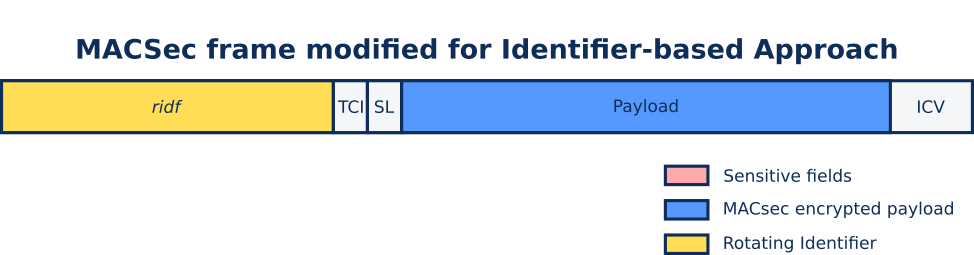}
  \caption{}
  \label{fig:macsec_mod_idf}
  \end{center}
  \end{subfigure}
  \begin{subfigure}{\columnwidth}
  \begin{center}
  \includegraphics[scale=1.0,keepaspectratio=true]{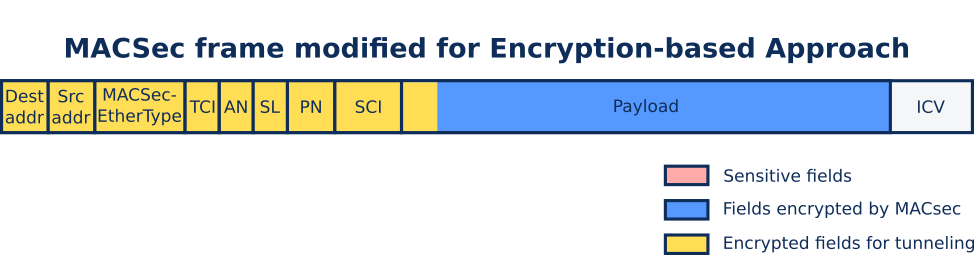}
  \caption{}
  \label{fig:macsec_mod_enc}
  \end{center}
  \end{subfigure}
  \caption{Standard and proposed MACsec frame structures. Field lengths are not to scale.}
\end{figure}

Both approaches build on the ``naive'' approach discussed in the introduction, where a simple and insecure tunneling protocol was used.
Hence, in the following, we will assume an underlying protocol, that actually facilitates the tunneling, meaning a Layer 2 Ethernet frame is repackaged into Layer 3 IP packets and sent to the remote tunneling gateway.
All modifications discussed below assume, that the resulting modified frame is transmitted over the untrusted network using some unprotected tunneling protocol, like L2TP or VXLAN.

The problem with the naive approach is, that security sensitive MACsec header information is sent in plain text.
The identifier-based approach wants to solve this by replacing these parts of the headers by a random identifier.
On first sight, the identifier would not reveal any information to a potential attacker and at the same time would not require any costly cryptographic operations.
Simple table lookups suffice at the uplink (sending gateway) and downlink (receiving gateway) and the necessary mappings between identifier and header contents can be exchanged via the management channel. 

We classify frames into flows according to the concept introduced above. This results in most fields being constant inside of a flow class. Subsequently, every flow can be represented by one identifier, effectively masking and compressing these fields.
The PN is a field that is security-critical and which changes on each frame, but since it is predictable, there is no need to send it in each frame.
The TCI and SL fields may change, but as discussed above, they are not critical towards security and hence can just be transmitted without additional measures.
Fig.~\ref{fig:macsec_sensitive} gives an overview on the MACsec header fields that are deemed critical.

A tunnel gateway needs to register a new flow when it occurs and share that information with the remote gateways.
Then each time a frame of a respective flow needs to be tunneled, the sensitive header fields are removed and the identifier is added instead.
This can be done very fast and efficient and hence results in only a small overhead for the tunneling operation.
All valuable information is either removed or masked (the payload is already encrypted by the MACsec devices).

Yet, the result of these first considerations is still insecure.
The core problem is, that no strong form of authentication is included.
All packets of the same flow share the same identifier, making it easy for an attacker to forge valid traffic as it is trivial to predict a valid identifier after having observed one, opening the possibility of DoS attacks. This also makes it easy for an attacker to link packets to flows, which can be considered a confidentiality violation.
Additionally, this approach does not assume packets getting lost or blocked in transit.
The receiving gateway can not know if a packet is missing and therefore might reconstruct the wrong PN.

Hence, we improve on this idea by using rotating identifiers and by adding a receive window.
Now, flows are still bound to a base identifier \textit{bidf}, that is also sent to the remote gateways via the management channel. 
Yet, this identifier is not used directly to replace the header fields.
Instead, it is used to derive a rotating identifier \textit{ridf} using a secure derivation function $F$ (\eg a hashing function) that takes the respective PN as input: $ridf_{PN} = F \left(bidf, PN\right)$.
The resulting \textit{ridf} is different for each tunneled frame and looks random for an attacker.
Additionally, all \textit{ridfs} can be calculated independently of each other as they are only based on the \textit{bidf} and the predictable PN.
$F$ should be a cryptographically strong function, meaning attackers should not be able to recover the original \textit{bidf}.
Further, it should be collision resistant, meaning it should be very improbable to arrive at the same \textit{ridf} using different inputs.
Fig.~\ref{fig:macsec_mod_idf} shows the resulting frame.

\newcommand{\mc}[3]{\multicolumn{#1}{#2}{#3}}

\begin{table*}
\caption{Tables necessary for management of Flows.}
\begin{subtable}{.30\textwidth}
\begin{center}
\begin{tabular}{|l|l|}
\mc{2}{l}{\textbf{Flow Table Entry Uplink}}\\
\hline
\textbf{Key} & SCI | SA\\
\hline
\textbf{Values} & - Unicast bidf\\
 & - Broadcast bidf\\
 & - Remote Gateways\\
 & - Timeout\\
\hline
\end{tabular}
\caption{Uplink.}
\end{center}
\end{subtable}
\begin{subtable}{.30\textwidth}
\begin{center}
\begin{tabular}{|l|l|}
\mc{2}{l}{\textbf{Flow Table Entry Downlink}}\\
\hline
\textbf{Key} & bidf\\
\hline
\textbf{Values} & - Header Data\\
 & - Window\\
 & - Next expected PN\\
 & - Bound\\
\hline
\end{tabular}
\caption{Downlink.}
\end{center}
\end{subtable}
\begin{subtable}{.30\textwidth}
\begin{center}
\begin{tabular}{|l|l|}
\mc{2}{l}{\textbf{Identifier Table Entry}}\\
\hline
\textbf{Key} & ridf\\
\hline
\textbf{Values} & - PN\\
 & - Seen (Replay detection)\\
 & - Pointer to Flow\\
 & \hspace{2mm}Table Entry\\
\hline
\end{tabular}
\caption{Downlink.}
\end{center}
\end{subtable}
\label{tab:flow_tables}
\end{table*}

Replacing sensitive header fields with rotating identifiers makes the protocol more complicated.
The flows are managed in tables (Tab.~\ref{tab:flow_tables}).

The uplink procedure (putting an incoming frame into the tunnel) is the following:
the flow, an incoming frame belongs to, is checked based on the SCI, SA and destination Ethernet address and if a new flow is identified, a \textit{bidf} is generated.
This base identifier is a simple random number and is sent via the management channel together with the MACsec header fields, it is supposed to replace, to all remote gateways.
This includes the PN.
Then an appropriate \textit{ridf} is calculated and then used to replace the respective MACsec header fields in the frame.
The new frame is then sent to the remote gateway.

The downlink procedure includes additional steps.
First, when a tunnel gateway receives information about a new flow, it adds that flow to the so-called Flow Table.
This table is a key-value list, where the key is the \textit{bidf} and the value consists of other necessary state information, like header values etc.
Additionally, the gateway will use the received PN to precalculate a window of next expected possible \textit{ridf}, based on the received \textit{bidf} and PN. These are added to the so-called Identifier Table, where the key is the \textit{ridf} and the value is further information, like the PN. One of those values is a pointer back to the respective Flow Table entry, the \textit{ridf} belongs to.

When a packet arrives from the tunnel, the \textit{ridf} found inside the frame is looked up in the Identifier Table.
If the identifier is found, the PN entry and the pointer into the Flow Table is used to reconstruct the original frame.
Additionally, the old Identifier Table entry is removed and a new one is calculated and added. The next expected identifier is updated as well.

We took the approach of allowing for multiple possible next \textit{ridfs} to account for packet loss inside a defined sliding window.
The Identifier Table is managed so that it always holds the appropriate \textit{ridfs} corresponding to the sliding window.

As described above, tunnel gateways are initially oblivious to behind which remote tunnel gateway the destination of a new flow is.
We account for that with a two step process.
First, incoming frames from new flows are broadcasted to all remote gateways.
When the destination MACsec device answers, that triggers the flow discovery process on the remote tunnel gateway.
As a result the first gateway learns the destination of the new flow.
It registers this information in its Flow Table and sends future frames from that flow only in this direction.

MKA automatically establishes one-to-one SCs with every participant.
Consequently, a broadcast message is then sent multiple times over each SC.
This behavior results in a problem for our flow-based design.
Independently of whether a frame is sent to a specific destination address or to the Ethernet broadcast address, the same PN is increased at the sender's SA.
Yet, our tunneling gateways would see different destination addresses (unicast or broadcast address) and would hence attribute this frame to different flows.
The increase of the PNs would only be registered at one of those flows.
As a result a wrong PNs would be included in subsequent reconstructed frame.

To confront this issue, we need to bind both flows together.
This is done by registering both flows separately on the tunnel gateways, but by updating the PN as well as the sliding window, when a new frame arrives also on the respective other flow to keep both synchronous.

Finally, we will discuss the fact, that the \textit{ridfs} can be guessed by an attacker.
If that occurs, a resulting properly reconstructed false frame will still be dropped by the MACsec device in the local network, because of MACsec's integrity check.
This means that the only possible type of attack on the local device, is a DoS attack.
The success of such an attack depends on the possible rate with witch an attacker could guess correctly. With a sufficient length of the \textit{ridf}, such an attack becomes very inefficient and hence improbable.
In general terms, using random identifiers like our \textit{ridfs} is comparable to using cookies for DoS attack mitigation, like for example Wireguard does. This is an established approach and considered a good defence against this type of attack.

\subsection{Encryption-based Approach}

The previous approach tried to improve performance by avoiding cryptographic operations like encryption and integrity checks and replacing them with a message-independent calculable single derivation function.
Yet, it requires additional steps for flow attribution and binding, which increases complexity of the protocol. 

The approach discussed in this section on the other hand will be much simpler by relying on encryption of the MACsec headers.
We still assume a secure management channel between tunnel gateways.
This allows to use symmetric encryption and we assume appropriate key material has been exchanged when the tunnel gateways were set up.

The sensitive header fields that need to be encrypted are source and destination address, PN, SCI and AN.
These amount to 194 bits.
As we chose the block cipher AES for encryption and as it has a block size of 128 bits, this results in two cipher blocks of overall 256 bits length.
This is more space than necessary and in fact allows to bluntly encrypt the first 256 bits of the frame. This includes all MACsec headers and 32 bits of the payload segment, as is depicted in Fig.~\ref{fig:macsec_mod_enc}.

We encrypt the second block first and then use its plaintext as well as the resulting ciphertext as additional input for the encryption of the first block. This is corresponds to the PCBC (Propagating Cipher Block Chaining) mode of operation. We chose this mode in order to provide integrity and confidentiality without additional initialization vectors or integrity check values. This reduces the frame size and saves expensive cryptographic operations. This is possible due to the unique properties of our approach.

As described, the second block includes 32 bits of payload that is already encrypted by MACSec.
These 32 bits are always unpredictable for the attacker.
MACsec ensures this by using a different initialization vector (the PN) for each frame.
These 32 bits are enough to make our second block unpredictable as well.
This is due to the diffusion property of AES (which we use for encrypting our blocks).
It means, that a single bit change in the plaintext results in a completely different ciphertext.

Consequently, the ciphertext of the first block also becomes unpredictible through the mode of operation though it could contain identical plaintexts across multiple MACsec messages, as we use the encrypted second block to construct it.
As a result, both blocks of ciphertext are different for every frame, independently of the repeating header fields.
The only requirement hereby is, that the symmetric encryption keys have to change, when any of the tunneled SAs changes, because then the ciphertext of the MACsec payload might repeat.

We assume the gateways to maintain Flow Tables similar to the identifier-based approach above.
Hence, the tunnel gateway can send the frame to the remote gateway, which then decrypts it with previously exchanged keys.
The remote gateway looks up the header information in the Flow Table and if the entry matches to the decryption result, the frame is put onto the internal network. Otherwise it is discarded.

Finally, we again do not explicitly add integrity protection to our encrypted headers.
Instead, the lookup in the Flow Table facilitates that, as a single bit flip during transit would result in a completely different decrypted plaintext.
The attacker is not able to generate ciphertext which decrypts to valid plaintext (i.e. resulting in a successful lookup), without knowing the secret key.

\section{Implementation}
\label{sec:implementation}

\begin{figure}
\begin{center}
\includegraphics[width=\columnwidth,keepaspectratio=true]{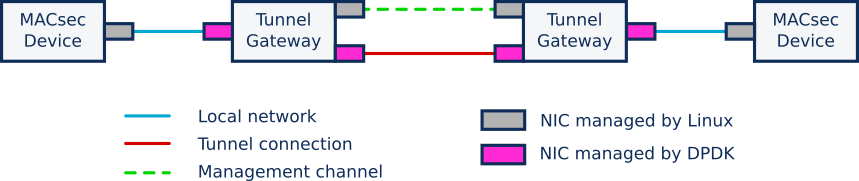}
\caption{Setup used for implementation and evaluation.}
\label{fig:test_setup}
\end{center}
\end{figure}

The setup used to test and evaluate our implementation is shown in Fig.~\ref{fig:test_setup}.
All involved NICs are Gbit Ethernet devices.
The tunnel gateways were implemented using Netgate MBT-4220 appliances.
They are equipped with an Intel Atom E3845 CPU with 1.91 GHz, 2 GB RAM and two on-board Intel I210 Gigabit NICs.
The management channel was realised by USB-Ethernet adapters attached to the tunnel gateways.
These interfaces are rather slow due to the USB interface, but as they only processed management traffic, the tunneling traffic, we actually wanted to optimize, was not affected.
The tunnel traffic was exclusively transmitted through the on-board network interfaces.
Both gateways ran CentOS Linux 8 with a kernel version of 4.18.
The MACsec devices were implemented using ODROID-H2 minicomputers from Hardkernel.
These come with a quad-core Intel Celeron J4105 with 1.50 GHz and ran Ubuntu 20.04.1 LTS with a Linux kernel in version 5.8.
The performance-critical network interfaces in pink were bound to DPDK, in effect hiding them from the Linux kernel.
No other system service had access to these interfaces and could interfere with the measurements.
We used the igb\_uio kernel module driver to manage the NICs.

We required a TCP stack for the Management channel for reliable transmission, but existing implementations on top of DPDK did not prove to be flexible enough.
Instead, we decided to use DPDK for the tunnel connection only and add an additional interface through a USB-Ethernet adapter for the management channel.
It was not bound to DPDK and therefore allowed to be managed by the Linux kernel. We used Wireguard as security layer. 

We used DPDK in version 19.11.2, OpenSSL in version 1.1.1g and libsodium in version 1.0.18.
Our DPDK application was compiled using GCC in version 8.3.1 and we always used the compiler flags, recommended by the DPDK manual.
Note that our setup is a reduced version of the scenario introduced above.
Yet, we of course implemented the steps of the protocol that are concerned with remote gateway management and selection.

To implement our tunneling approaches, we created a MACsec parsing library using DPDK that allowed for extracting information from incoming frames and to construct outgoing ones.
To actually transport the frames, we also implemented a prototypical tunneling protocol in DPDK that mirrors the functionality of \eg VXLAN.
The Flow and Identifier Tables described in Tab.~\ref{tab:flow_tables} were implemented using DPDK hash tables.
While MACsec frames that transported payload were transmitted using the tunnel connection, MKA (management) traffic was transmitted over the management channel.
MKA sends synchronization frames only every second and hence does not strain the channel much.
Additionally, MKA traffic is generally not performance-critical.

The derivation function $F$ is the central cryptographic primitive of the identifier-based approach.
Hence, it had to be chosen with special care.
The main requirement was that it must be keyed, meaning an additional value can be inserted to make it nondeterministic.
Function classes that could fulfill this requirement, are encryption and hashing functions.
We chose SipHash \cite{siphash}.
It describes a family of hashing functions that is optimized for performance and short inputs and has been designed for use in hash tables and message authentication codes.
This corresponds perfectly to our use case and can be considered state-of-the-art.
We use SipHash-2-4 as these parameters provide maximum security according to the authors.
For input, we use a base identifier (\textit{bidf}) of 128 bit length, with the 32 bit PN as hash key.
The output is our rotating identifier \textit{ridf} with a length of 64 bit.

The encryption-based approach was simpler to implement compared to the identifier-based one.
For flow discovery, flow management and lookup of remote gateways, we used the same DPDK hash table implementation from above.
The header encryption was implemented using AES provided by the OpenSSL library.
The decryption step at the remote tunnel gateway was implemented equally straight forward.
It includes a lookup to verify that the frame belongs to a valid flow.

During all experimental runs, MACsec was used in encryption mode and it was made sure that, although handling was implemented into the protocol, no MKA traffic happened during the experiments.

\section{Evaluation}
\label{sec:evaluation}

This section will first discuss our proposed protocol designs and then present results of performance measurements done on our experimental implementation.

\subsection{Protocol Designs}

The main goal of the proposed approaches was to ensure the confidentiality of header information when MACsec frames are being tunneled over insecure networks.
Both approaches accomplish that. The encryption-based approach by simply encrypting the whole header and the identifier-based approach by replacing the sensitive fields with a random identifier.

Timing information from traffic patterns are still observable.
Yet, this is out of scope as it requires different countermeasures like buffering and sending in regular intervals. These measures would also destroy the performance.

We did not consider the possibility of attacks from the inside, meaning an attacker that sits in one of the local networks.
He could for example inject random MACsec frames that would each time create a flow entry on the tunnel gateway.
This would eventually lead to memory exhaustion, because the gateways in our scenario do not know which flows are benign.
Yet, there are countermeasures available, that could be added to our protocols.
For example, if MKA was available, gateways could analyze the MKA traffic and derive the benign flows.
Another possible approach could be quotas for SCIs or source Ethernet addresses or shorter timeout times, when no remote gateway signals a successful answer to that frame.

Performance-wise, the approaches mainly differ in the cryptographic operations used.
For each transport of a frame within the identifier-based approach, 3 times the cryptographic hash function has to be called. Once on the uplink to create the \textit{ridf} and two times on the downlink to adjust the sliding window by calculating new expected \textit{ridfs}.
The encryption-based approach on the other hand needs to encrypt and decrypt two AES blocks.
The first approach indeed works with the fewest amount of cryptographic operations possible and through the header replacement it even reduces the effective size of the packets, allowing for bigger payloads.
Therefore, on paper, there should be some difference between the performance results. 

The other operation which might impact performance, is the Flow Table lookup.
Both approaches used DPDK hash tables for that purpose.
A lookup takes constant time, independently of the table's size, assuming enough memory is available for them.
In our experiments this was always the case, no swapping happened.
  
In any case, some kind of table lookup is not avoidable anyhow if the scenario considers more than two gateways, as the destination gateway has to be looked up, and considering that broadcasting to all remote gateways is not an option.
Furthermore, the lookups realize authorization of incoming traffic on the downlink and provide DoS protection, as we described above.

\subsection{Performance}

\begin{table}
\begin{center}
\caption{Performance measurements including standard deviation (mdev) of regular experimental setup.}
\label{tab:perf_norm}
\begin{tabularx}{\columnwidth}{|c|X|r|r|}
\hline
\textbf{\#} & \textbf{Scenario} & \textbf{RTT $\pm$ mdev} & \textbf{Throughput $\pm$ mdev}\\
\hline
1 & VXLAN & 3 $\pm$ 0.13 ms& 837 $\pm$ 9 Mbit/s\\
2 & VXLAN + Wireguard & 4.85 $\pm$ 0.55 ms & 245 $\pm$ 16 Mbit/s\\
3 & Identifier-based & 1.3 $\pm$ 0.07 ms& 842 $\pm$ 3 Mbit/s\\
4 & Encryption-based & 1.3 $\pm$ 0.05 ms& 837 $\pm$ 3 Mbit/s\\
\hline
\end{tabularx}
\end{center}
\end{table}

All of the measurements described in the following were conducted on the setup depicted in Fig.~\ref{fig:test_setup}.
The measurements were taken after initializing the connection with a simple ping.
This includes the establishment of flows at the gateways as well as necessary ARP resolution at the MACsec devices.

We measured different scenarios.
First, we bridged MACsec frames using standard VXLAN.
No protection of MACsec headers occurs, as VXLAN just repackages frames and transmits them using UDP.
This state-of-the-art approach should incur the least amount of overhead and the measurement results of this scenario shall serve as an indicator of what performance is actually achievable.
In the second scenario, MACsec frames were tunneled using a standard VPN protocol.
This is the state-of-the-art approach to solve the problem this work is based on.
We implemented the tunnel in this scenario using Wireguard, because it is considered highly efficient.
Finally, we also measured both the Identifier-based as well as the Encryption-based approaches in scenarios three and four.

For each scenario, we measured the latency as well as the throughput of the tunnel between the two MACsec devices.
Results are listed in Tab.~\ref{tab:perf_norm}.
The latency was measured as the mean of 65,535 individual ping round trip times (RTT) with a frame size of 64 bytes.
The throughput was measured using the tool iperf3 (using an interval of 10 seconds). 

The results of the first two scenarios show expected behavior.
The tunneling gateways are in principle capable of fully saturating the line speed of the 1 Gbit/s line (scenario one).
As the whole data stream had to be additionally encrypted, scenario two showed a huge decrease in performance. While latency was not so much affected, the throughput was reduced to one third.

The results of our approaches showed superior behavior compared to the first two.
Round trip times were reduced by half, even compared to the scenario without additional protection measures. Both approaches also managed to saturate the line.
Yet, differences between our approaches were minimal and performance-wise we could not decide on a winner.

General learnings from the results are, that latency can be greatly improved by circumventing the Linux network stack and that cryptographic operations have big influence on the performance if applied to whole frames.
Offloading functionality to DPDK improved the (latency) performance considerably.
The results show, that our approaches greatly reduce the overheads when tunneling MACsec frames compared to the state-of-the-art.

\begin{figure*}
\begin{subfigure}{\columnwidth}
\begin{center}
\includegraphics[scale=0.45]{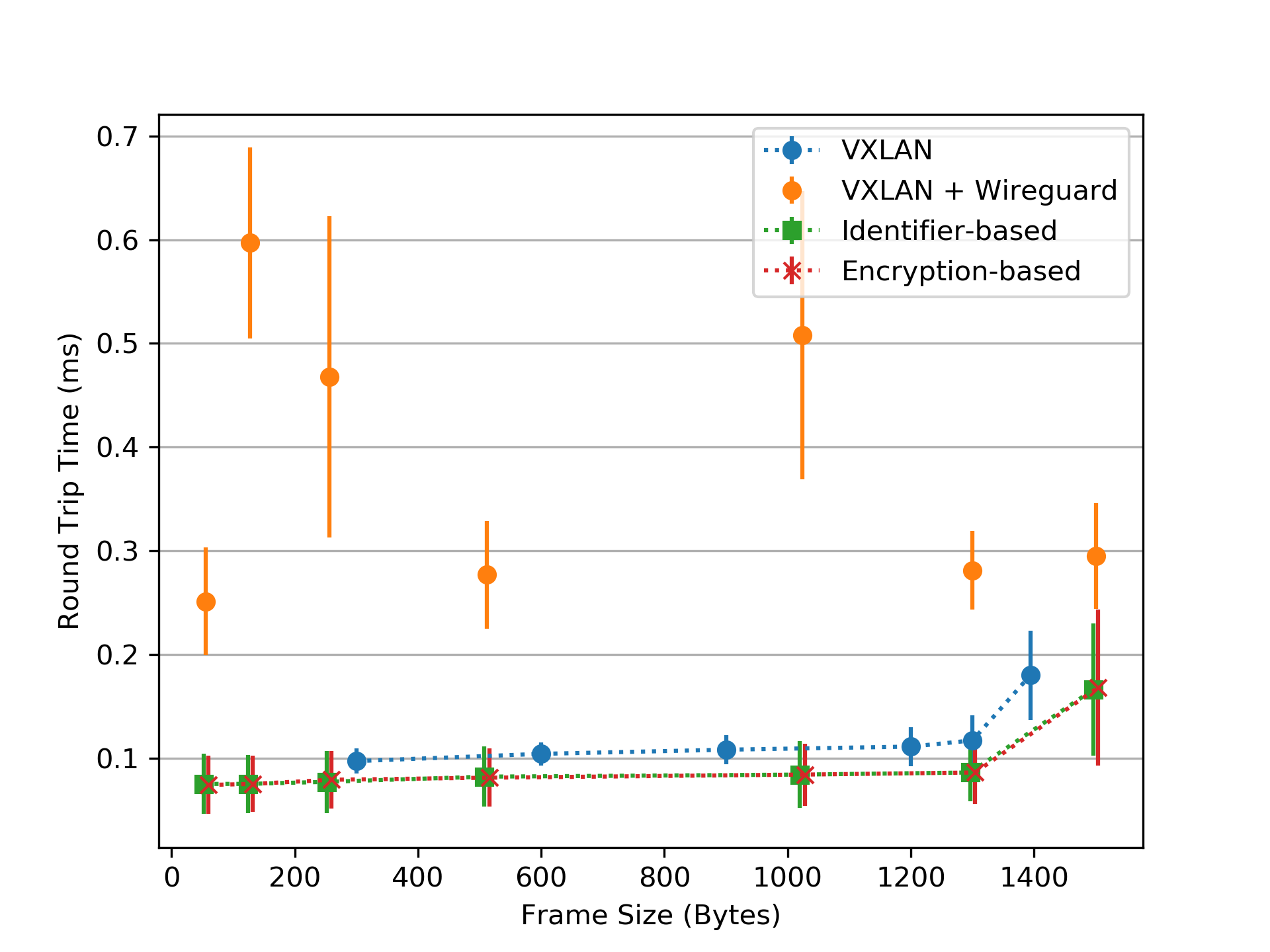}
\caption{Round trip times with standard deviation.}
\label{fig:latency}
\end{center}
\end{subfigure}
\begin{subfigure}{\columnwidth}
\begin{center}
\includegraphics[scale=0.45]{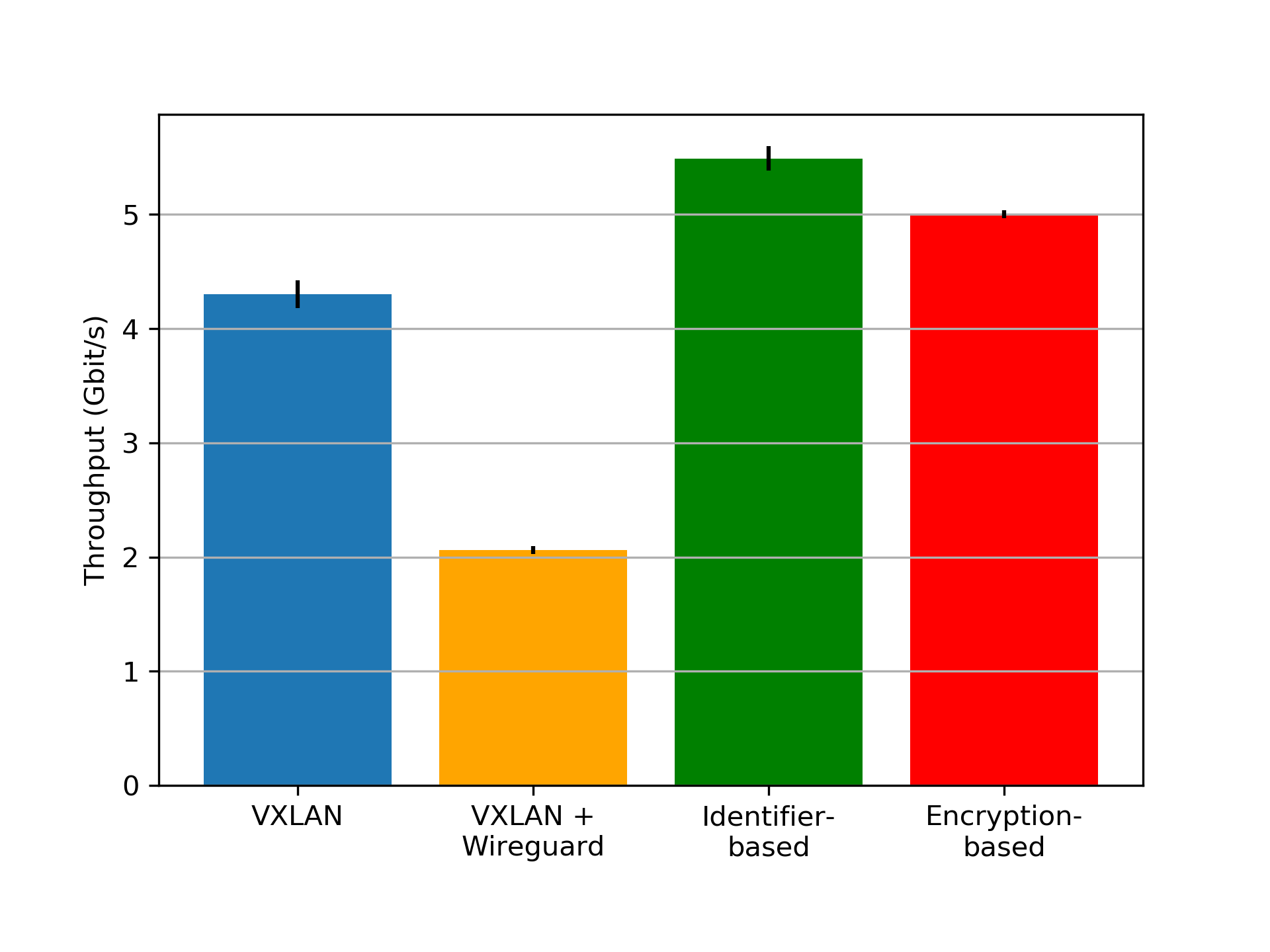}
\caption{Throughput with standard deviation.}
\label{fig:bandwidth}
\end{center}
\end{subfigure}
\caption{Performance measurements of additional experimental setup.}
\label{fig:perf}
\end{figure*}

Yet by how much exactly, we could not determine, as the bottleneck in our measurement setup was not, as expected, the CPUs of the tunnel gateways but rather the Gigabit Ethernet interfaces.
Therefore, we measured all scenarios a second time using different hardware for the tunnel gateways.
This time we used two server platforms that were equipped with 16-Core Intel Xeon CPUs at 3 GHz, 64 GB RAM and four 10 Gigabit Ethernet NICs on-board.
Fig.~\ref{fig:perf} shows the results. Experiments were conducted in the same fashion as before.
We additionally measured the latency for different frame sizes, to see how that parameter might affect the results.
The jump at high sizes happens when the input frame size exceeds the maximum payload size, resulting in a packet split. 

Scenarios one and two show mostly expected behavior. 
Additional encryption incurs immense overheads. The throughput gets halved, while the latency shows erratic behavior for scenario two.
This probably stems from the fact that on top of the Linux networking stack two nested protocols are running.
A single frame gets queued and dequeued many times during transit.
Frame size clearly has no influence on the latency results and the unpredictable behavior may stem from cache misalignments, yet we cannot be certain.

Our approaches in scenario three and four on the other hand show superior performance.
Both compare vastly better than the state-of-the-art approach of scenario two and even better by some margin compared to the insecure scenario one.
And while both approaches show same latency behavior, the identifier-based approach achieves 10\% more throughput compared to the encryption-based approach.
This stems from the different header lengths of the approaches. The identifier-based approach effectively shortened the header by replacing the fields with the shorter rotating identifier.
Whether the performance margin is enough to justify the increased complexity of the approach is probably dependent on the ultimate use case and cannot be decided here.

\section{Conclusion}
\label{sec:conclusion}

This work investigated two approaches how MACsec frames can be tunneled in a secure and efficient way. One approach was optimized for performance as much as possible, while the second could be implemented much simpler.
Results showed both approaches to be significantly better compared to the state-of-the-art.

Our work offers many pointers for future research.
A non DPDK version running inside the Linux kernel might yield acceptable performance and at the same time increase the applicability of our approach and allow for better comparison.
MKA traffic could also be handled differently. We piped it over the management channel as it was the easiest way, yet it could also be transferred over the tunneling channel.
Further integration could also lead to detection of attackers inside the local network.
This scenario is excluded so far.
Furthermore, certain hardware acceleration features could be researched to increase performance even more. This includes leveraging CPU parallelism, where DPDK provides much potential.
We used only one core so far.
Finally, even more sophisticated acceleration technologies could be investigated, like for example FPGA-powered SmartNICs.

\begin{acks}
This work was co-funded by SAB (Development Bank of Saxony) under frameworks from ERDF (European Regional Development Fund) and ESF (European Social Fund), by public funding of the state of Saxony/Germany and by the German Research Foundation (DFG) as part of Germany’s Excellence Strategy EXC 2050/1 – Project ID 390696704 – Cluster of Excellence Centre for Tactile Internet with Human-in-the-Loop (CeTI) of TU Dresden.
\end{acks}

\end{document}